\documentstyle[11pt,epsf,paspconf]{article}
\begin{document}
\title{HI velocity fields and the shapes of dark matter halos}
\author{R.H.M.~Schoenmakers}
\affil{Kapteyn Astronomical Institute,  
         P.O.~Box 800, 9700 AV, Groningen, The Netherlands.
         }
\begin{abstract}
  I report on a new method for measuring deviations from
  axisymmetry in the velocity fields of gas disks. The method
  is an extension of the method that Franx, van Gorkom and
  de Zeeuw developed for a single orbit gas ring. 
  The measurement is based upon a higher order harmonic expansion of
  the full velocity field. Epicyclic theory is used to calculate the
  effects of a perturbation in the potential on these harmonic
  terms. It is shown how the $s_1$ and $s_3$ harmonics can be used to
  measure $\epsilon_R \sin(2\phi_{obs})$, where $\epsilon_R$ is the
  elongation of the potential and $\phi_{obs}$ is an (unknown) viewing
  angle. The advantage of this method over
  previous attempts to derive the elongation of dark matter halos is
  that, by using HI, we can probe the potential at radii beyond 
  the stellar disk, into the regime where the dark matter is thought to
  be the dominant dynamical component.

  As a demonstration, I applied this method to HI velocity fields of two
  spiral galaxies, NGC~2403 and NGC~3198. NGC~3198 shows a very small
  $\epsilon_R \sin(2\phi_{obs})$, which suggests that the galaxy is very
  nearly axisymmetric. NGC~2403 shows a larger $\epsilon_R
  \sin(2\phi_{obs})$, varying in a systematic way between $0$ and
  $0.1$ and is probably influenced by spiral arms. The current data suggests
  that spiral galaxies are close to axisymmetry, but a larger sample is
  needed to quantify this statement.
\end{abstract}

\section{Introduction}

It is generally accepted that disk galaxies have massive dark halos,
but little is known about the shape of the dark matter halos.  Dark
halos were generally modeled as being spherical until Binney (1978)
argued that the natural shape of dark halos is triaxial. If halos are
indeed triaxial, the disk of the galaxy will reside in one of the
principal planes of the dark halo.  In a triaxial halo the axes obey
the relation $a > b > c$. I will assume that the disks of spiral
galaxies lie within the $a,b$ plane.

Triaxial dark matter halos occur naturally in cosmological N-body
simulations of structure formation in the universe (Katz \& Gunn 1991,
Dubinski \& Carlberg 1991, Dubinski 1994), but the exact distribution
of the halo shapes is as yet uncertain. Especially the inclusion of
baryons in the simulations tends to make the halos more oblate
(Dubinski 1994). So the observed distribution of
the shapes of dark matter halos will be a powerful constraint on
scenarios of galaxy formation and subsequent evolution.

Measuring the shapes of dark halos can be split into two parts:
measurement of the ratio $c/a$, i.e. the flattening perpendicular to the
plane of the 
disk, and measurement of the intermediate to major axis ratio
$b/a$, i.e. the elongation in the plane of the disk.
The axis ratio $c/a$ can be measured using, for instance, polar ring
galaxies (e.g. Whitmore, McElroy \& Schweizer 1987, Sackett \& Sparke
1990, Sackett {\it et al} 1994), kinematics of halo stars of the Milky
Way (van der Marel 1991) or the flaring of HI-disks (Olling, these
proceedings, Sicking 1996). All these measurements seem to indicate
that $c/a < 1$, in other words, a spherical halo is excluded by these
measurements.  

The elongation of the dark halo ($b/a$) is a similarly uncertain
parameter. Attempts to measure this quantity include studies of the
inclination distributions of large samples of spiral galaxies and
fitting models with different elongations to them (Lambas 1992, Binney
\& de Vaucouleurs 1981, Fasano {\it et al} 1992), detailed
observations of the old stellar disks of spiral galaxies (Rix \&
Zaritsky 1995), kinematics of the milky way (Kuijken \& Tremaine
1994) and the scatter in the Tully--Fisher relation (Franx \& de
Zeeuw 1992). All these measurements indicate that $b/a\sim 0.9$.

The only direct measurement of the elongation of the potential of a
disk galaxy so far is of the S0 galaxy IC~2006 (Franx, van Gorkom
\& de Zeeuw 1994, hereafter FvGdZ),
which contains a large HI ring in the plane of the disk. Using
information on both the kinematics and the geometry of the ring, FvGdZ
were able to measure the elongation of the potential at the location of
the ring. The measured elongation was consistent with zero. 
The method used by FvGdZ was based on the
assumption that one could use epicycle theory to predict the geometry and
velocity field of the ring in a mildly perturbed potential. If the
external potential is elongated, the velocity variation along the
ring will not be precisely sinusoidal, but higher harmonic terms will
be superimposed on it. Making a
harmonic expansion of the velocity field of the ring and interpreting the
measured harmonics within the framework of epicycle theory, given the
ring geometry, the
elongation of the potential at the position of the ring can be found. 

I have extended the FvGdZ formalism for measuring the elongation of
the potential of a single orbit gas ring to the case of a slightly
non-axisymmetric gas disk, which may contain spiral-like
perturbations. This analysis assumes a stationary perturbation and
closed stable orbits. Therefore, applicability to non-linear phenomena
like spiral arms is limited. But a small global elongation of the
overall potential, as is the case with a triaxial dark matter halo,
can be analysed with this method. The advantage of this method over
previous attempts to derive the elongation of dark matter halos is
that, by using HI, we can probe the potential at radii beyond 
the stellar disk, into the regime where the dark matter is thought to
be the dominant dynamical component.

\section{Results from perturbation theory}

Suppose the (cold) HI-disk resides in a potential
\begin{equation}
\label{potentiaal2}
V(R,\phi) = V_0(R) + V_m(R)\cos(m\phi + \varphi_m(R)),
\end{equation}
where $R$ and $\phi$ are polar coordinates in a frame that rotates
with the perturbation, at a pattern speed $\Omega_{m,p}$. Here,
$V_0(R)$ is the unperturbed potential, $V_m(R)$ the amplitude of the
perturbation and $\varphi_m(R)$ its phase. The perturbation is assumed
to be stationary. One can calculate the
possible closed loop orbits of the gas in this potential (analogous to Binney \&
Tremaine, 1987, p. 146). Subsequently one can determine the velocity
field generated by these orbits. Now assume that this velocity field
is observed under viewing angles ($\phi_{obs},i$), where $\phi_{obs}$
is the angle between the line where $\phi=0$
and the projection of the line-of-sight onto the plane of the disk and $i$ is the
inclination of the disk. If one then introduces the azimuthal angle
$\psi= \phi - \phi_{obs} + \pi/2$ (the angle in the plane of the disk
that is zero on the line of nodes), it is possible to
show that the velocity field projected on the sky (hereafter ``line-of-sight
velocity field'' or ``l.o.s.\ velocity field'') has the following form (express the
line-of-sight velocity as a Fourier expansion $v_{los} =
\sum_n^{} c_n \cos n\psi + s_n \sin n\psi$):
\begin{eqnarray}
\label{lineofsight}
v_{los} \!\!&=&\!\! v_* \left[ c_1 \cos \psi + s_{m-1}\sin(m-1)\psi +
  c_{m-1}\cos(m-1)\psi\qquad \right. \nonumber \\ && \left. \qquad\qquad
  \ +\; s_{m+1}\sin(m+1)\psi + c_{m+1}\cos(m+1)\psi\right],
\end{eqnarray}
\noindent
with $ v_* = v_c \sin i$ ($v_c$ is the circular velocity) and coefficients $c_x,s_x$ that depend on $m$, $\phi_{obs}$, $\varphi(R)$, $V_0(R)$, $V_m(R)$ and $\Omega_{m,p}$. 
Now the measurable parameters ($v_*, c_x, s_x$) are expressed in terms of
internal parameters for a potential perturbed by a single Fourier
term.

From equation (\ref{lineofsight}) we can conclude that {\it if the 
potential has a perturbation of Fourier number $m$, the l.o.s.\ velocity
field contains $m-1$ and $m+1$ Fourier terms}. Qualitatively, this
conclusion was also inferred by Canzian (1993). 

In order to measure the harmonic terms in the l.o.s.\ velocity field, we first
fit a set of tilted-rings to the l.o.s.\ velocity field using standard
tilted-ring fitting routines. Then the velocity along each ring is
decomposed into its harmonics.
In general, one does not a priori know the inclination $i$, position
angle $\Gamma$
and centre ($x_0,y_0$) of a galaxy. Since the l.o.s.\ velocity field itself is used to
determine these parameters, it is clear that this can influence the
resulting harmonic terms. FvGdZ showed that this fitting of the ring
parameters from the l.o.s velocity field will affect the expansion given in
equation (\ref{lineofsight}) due to a possible difference between the best fitting
parameters ($i$ and $\Gamma$) and the true parameters. It can be shown 
that only for $m=2$ there will
be first order differences between the best fitting and true internal
parameters, whereas for $m \neq 2$, equation
(\ref{lineofsight}) predicts the correct
form of the l.o.s.\ velocity field to second order.
In the $m=2$ case, a misfit of the kinematic centre will result in
additional $c_0,s_2$ and $c_2$ terms in the harmonic
expansion and misfitting of the inclination or position angle will affect
the $c_1,s_1$ and $c_3,s_3$ terms. 

In the case of a globally elongated potential, as caused by for
instance a
triaxial halo, we have $m=2$, $\Omega_{2,p} \approx 0$ and
$\varphi_2(R) \sim \mbox{const}$. Note that here we are in the $m=2$
case, where the best fitting $i$ and $\Gamma$ are not equal to the
true $i$ and $\Gamma$. Assuming a flat rotation curve, one
finds that after a tilted-ring fit the l.o.s.\ velocity field has the
following form (as derived earlier by FvGdZ):
\begin{equation}
\label{firstvlos}
v_{los} = \hat c_1 \cos\hat\psi + \hat s_1 \sin\hat\psi+ \hat c_3 \cos
3\hat\psi + \hat s_3 \sin 3\hat\psi,
\end{equation}
with
\begin{eqnarray}
\hat c_1 &=& v_c \sin i (1-{\textstyle{1\over 2}}\epsilon_v \cos 2\phi_{obs}),
\nonumber \\
   \hat s_1 &=& v_c \sin i\left[ 1-
    {{(3q^2+1)^2}\over{(3q^2+1)^2+(1-q^2)^2}}\right]
  \epsilon_R \sin 2\phi_{obs}, \nonumber \\ 
  \hat c_3 &=& 0, \\
  \hat s_3 &=&
  v_c \sin i \left[ -
    {{(1-q^2)(3q^2+1)}\over{(3q^2+1)^2+(1-q^2)^2}}\right]
  \epsilon_
R \sin 2\phi_{obs}, \nonumber
\end{eqnarray}

where $q \equiv \cos i$, $\hat\psi$ is the angle in the plane of the
ring ($\hat\psi$ is zero on the apparent major axis of the ring),
$\epsilon_R$ is equal to the elongation of the potential and \mbox{$\epsilon_v =
{2\epsilon_R}$}. We see that from the $\hat s_1$ and $\hat s_3$ terms
we are able to derive $\epsilon_R \sin(2\phi_{obs})$, the quantity
that I will try to measure.
Unfortunately, one cannot determine $\phi_{obs}$ separately, so the
combination of viewing angle and ellipticity is all one can determine
for an individual galaxy.
Global ellipticity will create $s_1$ and $s_3$ terms that are
{\it constant} with radius (and thus $\epsilon_R \sin(2\phi_{obs})$
will be constant as a function of radius as well), whereas spiral-like
perturbances
will cause $s_1$ and $s_3$ terms that change sign as a function of
radius and $\epsilon_R \sin(2\phi_{obs}) $ will wiggle.  
Therefore, I will interpret an average value of $\epsilon_R
\sin(2\phi_{obs})$ that is offset from zero as due to ellipticity of
the disk.  

Strong warping will invalidate the approach discussed above.

\section{Some example velocity fields}
\label{examfields}

In order to clarify the equations presented in section 2, we present \begin{figure}[t]
\label{figEXAMPLE}
\plotone{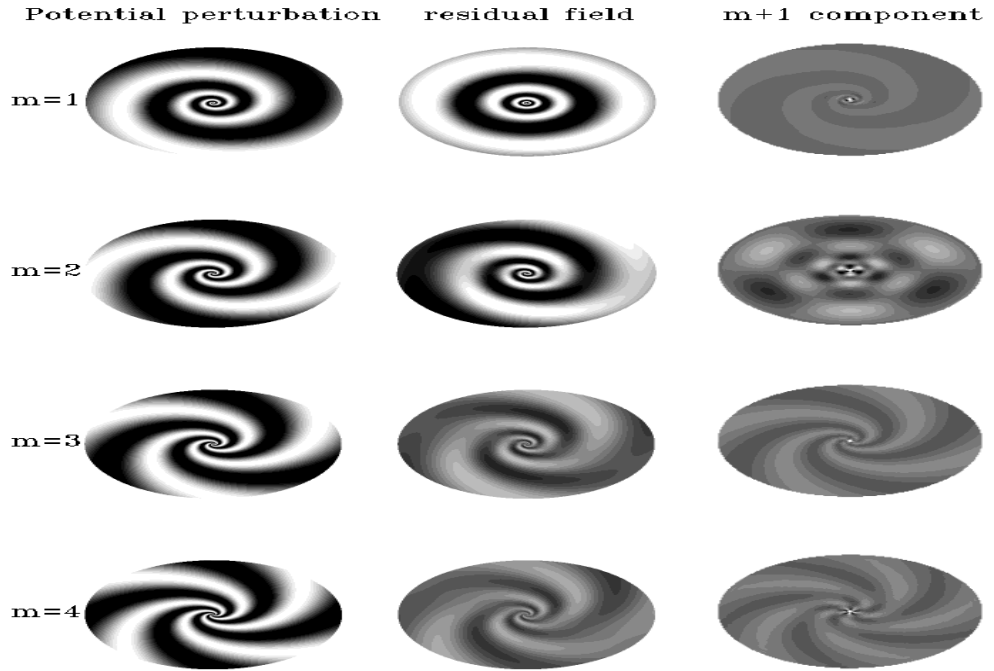}
\caption[]{Potential perturbations and velocity fields. We created a
  potential with a perturbation as given in equation (\ref{potentiaal2}) with
  a single $m$ (ranging from 1 to 4, shown in column 1).
 Subsequently we calculated the
  velocity field corresponding to this potential using equation (2) and made a
  tilted-ring fit to this field. This gave us a residual velocity
  field (column 2). This residual field is dominated by the $m-1$
  component. A harmonic fit to the velocity field revealed that the
  residual field contained the $m+1$ term
  as well (column 3).}
\end{figure} a number of example potential perturbations and corresponding residual
velocity fields in figure [1]. The harmonic terms of the different potential
perturbations range from $m=1$ to
$m=4$.  For the phase of the fields, $\varphi_m(R)$, we chose a
logarithmic spiral. The amplitude of the perturbation
has been taken constant throughout the field.
The l.o.s.\ velocity fields are then created using equation
(\ref{lineofsight}) (these are not shown).
A tilted-ring fit was then made to these l.o.s.\ velocity fields and
subsequently the harmonic fit was made,
revealing the two harmonic components that were hidden in them.

In figure [1] we see from left to right

\begin{enumerate}
\item{The $m$ term potential perturbation.}
\item{The residual velocity field (velocity field minus fitted circular
velocity) caused by this potential perturbation
(i.e. the $m-1$ plus $m+1$ terms together).}
\item{The $m+1$ component of the residual field.}
\end{enumerate}

We see that indeed an $m$ term in the density causes $m-1$ and $m+1$
terms in the kinematics of the gas. In these examples, the $m-1$ term
is dominant.  

Equation (2) predicts that the $c_3$ term should be fitted to zero by
the tilted-ring fit in the case of a $m=2$ perturbation in the
potential. Thus, in the $m+1$ component of the residual field, no
spiral-like structure should be visible any more. In figure [1], we see
that this is indeed the case. Instead of spiral structure, we see radially alternating positive and negative
contributions, caused solely by the $s_3(R)$ term.
If the potential would have contained a global intrinsic
ellipticity, $m=2, \varphi_2(R)=\mbox{const}$, the $s_1$ and $s_3$
terms caused by this elongation would have been constant as a function of radius. Therefore, the
measured $s_3$ would have been constant and non-zero as well and no
radially alternating pattern for $s_3$ would have been
visible in this plot. 
 
Since in the fitting
procedure the centre was kept fixed, we also see a two-armed spiral in
the $m+1$ residual field of the $m=1$ potential perturbation. If we would have taken the centre as a free parameter
in the fit, it would have drifted in such a way as to make the $c_2$
and $s_2$ terms disappear. 

\section{Two test cases: NGC~2403 and NGC~3198}

\subsection{Data description}
 \begin{figure}
\plotone{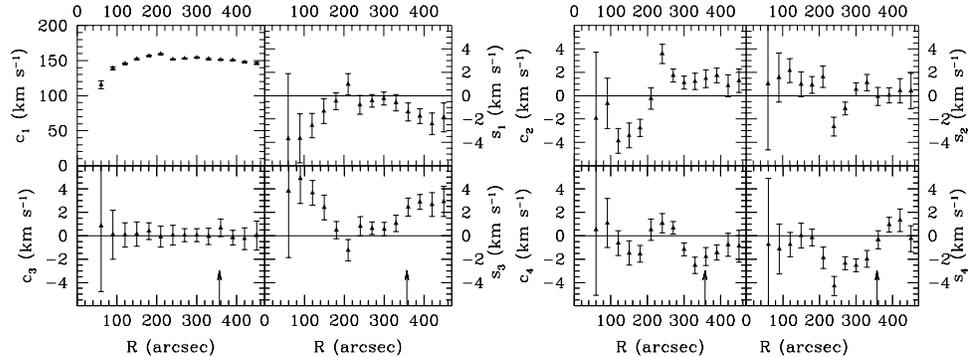}
\caption[]{Harmonic decomposition of the velocity field of
  NGC~3198. Only the first four terms of the expansion are shown, since
  higher order terms have low amplitudes. The $c_1$ term gives the circular velocity,
  i.e. the rotation curve. From the $s_1$ and $s_3$ terms
  $\epsilon_R(R) \sin(2\phi_{obs})$ will be measured. The $c_2,s_2$
  terms are zero if the kinematic centre of a ring is left free in the
  fit. Since the same (fixed) centre is taken for all rings here, the
  $c_2,s_2$ terms vary. Especially in the inner parts the $c_2$
  term is relatively large, indicating that NGC~3198 might be somewhat
  lop-sided. The $c_3$ term is zero if the inclination is fitted
  correctly (equation 4), which is obviously the case here. There is
  also some power in the $c_4,s_4$ modes. This may be caused by some
  three- or five-armed spiral structure (see figure 1). Visual
  inspection of the HI surface density map shows a three-fold
  spiral-like structure. The arrows denote the Holmberg radius.}
\end{figure} 
\begin{figure}
\plotone{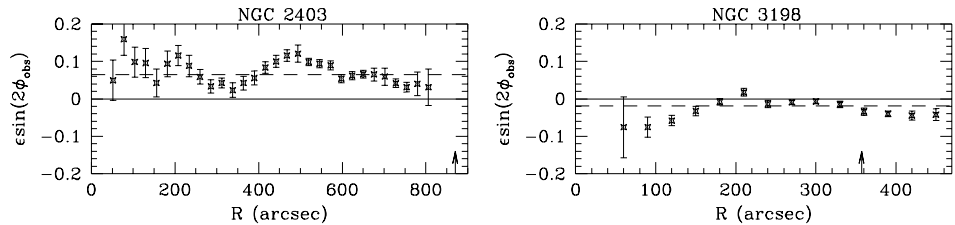}
\caption[]{ $\epsilon_R(R) \sin(2\phi_{obs})$ as a function of
  radius. The effect in NGC~2403 may be caused by spiral arms. The
  average value is used as an upper limit to the halo ellipticity.
  The measurement for NGC~3198 is relatively straight
  within errors and very low. The dotted line is the average value of
  $\epsilon_R(R) \sin(2\phi_{obs})$, the arrow denotes the Holmberg radius.}
\end{figure} 

NGC 2403 is a nearby (3.25 Mpc; Begeman, 1987) Sc(s)III galaxy
(Sandage \& Tammann, 1981). It has been observed with the WSRT, $4 \times
12$ hours (Sicking, 1996). The data have been smoothed
to a circular beam of 13 arcseconds.
The second relatively nearby (9.4 Mpc; Begeman, 1987) spiral galaxy we
will examine is the Sc(rs)I-II (Sandage \& Tammann, 1981) galaxy NGC
3198. This galaxy was observed by Sicking (1996) with the WSRT, also
for $4 \times 12$ hours. The observations have a
circular beam with FWHM of $18$ arcsec.
Both galaxies are well suited for our type of analysis: they have large
extents on the sky (Holmberg dimensions: NGC~2403: $29^{\prime} .0 \times
15^{\prime} .0$ and NGC~3198: $11^{\prime} .9
\times 4^{\prime} .9$), have a favourable inclination (NGC~2403: $i \approx
62^{o}$, NGC~3198: $i \approx 71^{o}$), the data have high
signal/noise, the beams are small and both galaxies show no obvious warp.
The residual maps of the galaxies show only small (typically $< 10$ km~s$^{-1}$) residuals, but they
appear to contain systemic structures.

\subsection{Measuring the elongation}

The first step in our procedure is to fit concentric tilted--rings to
the l.o.s.\ velocity fields of these galaxies. The width of
each ring is taken to be about twice the beam size, to be sure that all the
individual rings are independent.
Table [\ref{table1}] lists the radially averaged ring parameters found by the
tilted-ring fit.

\begin{table*}[hbtp]
  \begin{center}
    \leavevmode
\begin{tabular}{l|l|l}
Parameter & NGC~2403 & NGC~3198 \\ \hline
position angle &  $124\fdg 1 \pm 0\fdg 09$ & $215\fdg 93 \pm 0\fdg
13$\\
inclination &  $61\fdg 5 \pm 0\fdg 20$ & $70\fdg 6 \pm 0\fdg 21$ \\
systemic velocity &  $133.3 \pm 0.10$ km~s$^{-1}$ & $660.1 \pm 0.16$
km~s$^{-1}$ \\
$\epsilon_R \sin(2\phi_{obs})$ & $0.064 \pm 0.003$ & $-0.019 \pm
0.003$ \\
\hline
    \end{tabular}

    \caption{Radially averaged ring parameters for NGC~2403 and NGC~3198}
    \label{table1}
  \end{center}
\end{table*}

After the tilted--ring fit a harmonic fit was made to the velocity
fields along each individual ring. Harmonic terms were measured up to
ninth order. In figure [2], the result for the first four harmonics
of NGC~3198 are shown as an example.
The $c_3$ term is zero everywhere in both galaxies, indicating that
the inclination is correctly fitted for all rings. Furthermore, we
find that in both galaxies harmonics $c_n,s_n$ with $n>5$ are consistent with zero.
In the lower harmonics, clear systematic trends (as a function of radius)
are visible. Using $s_1(R)$ and $s_3(R)$, we are able to
measure $\epsilon_R(R) \sin(2\phi_{obs})$. The results of
these measurements
are given in figure [3], where we see that $\epsilon_R(R) \sin(2\phi_{obs})$ strongly
wiggles as a function of radius for NGC~2403, whereas for NGC~3198 it
is fairly constant within errors. Furthermore, note that
$\epsilon_R(R) \sin(2\phi_{obs}) > 0 $ for NGC~2403 at all radii. This
indicates that, although spiral arms are present that cause the
wiggling, there may also be a global elongation present, lifting 
$\epsilon_R(R) \sin(2\phi_{obs})$ from an average of $0$ to an average
of $0.064 \pm 0.003$. For NGC~3198 the average value of $\epsilon_R \sin(2\phi_{obs})$
is very low: $-0.019\pm 0.003$.

Since the effect of ellipticity and spiral arms cannot be separated
unambiguously, these measurements suggest that spiral
galaxies are close to axisymmetric. But more
galaxies are needed in order to quantify this result (Schoenmakers
{\it et al.}, in preparation).

\section*{acknowledgments}
I thank F.J.\ Sicking for stimulating
discussions and making available the data of both NGC~2403 
and NGC~3198. R.\ Bottema and T.S.\ van Albada are acknowledged for reading the
manuscript and discussions. I would like to thank M.\ Franx and P.T.\
de Zeeuw for their help and supervision of this project.

\vfill


\begin{references}
\reference  Begeman K., PhD-thesis, University of Groningen, 1987
\reference Binney J., 1978, \mnras, 183, 779
\reference  Binney J., Tremaine S., Galactic Dynamics , 1987, Princeton University Press
\reference Binney J., de Vaucouleurs G., 1981, \mnras, 194, 679
\reference Canzian B., 1993, \apj, 414, 487
\reference  Dubinski J., 1994, \apj, 431, 617 
\reference  Dubinski J., Carlberg R.G., 1991, \apj, 378, 496 
\reference Fasano G., Amico P., Bertola F., Vio R., Zeilinger W.W.,
1992, \mnras, 262, 109
\reference  Franx M., de Zeeuw P.T., 1992, \apj, 392, L47 
\reference  Franx M., van Gorkom J.M., de Zeeuw P.T., 1994, \apj, 436, 642 (FvGdZ)
\reference Gradshteyn I.S., Ryzhik I.M., 1965, Table of Integrals, Series
 and Products, New York: Academic Press  
\reference  Holmberg E., 1958, Medd. Lund Obs. Ser. II, 136
\reference  Katz N., Gunn J.E., 1991, \apj, 377, 365
\reference  Kuijken K., Tremaine S., 1994, \apj, 178, 421 
\reference Lambas D.G., Maddox S.J., Loveday J., 1992, \mnras, 258, 404
\reference  van der Marel R., 1991, \mnras, 248, 515
\reference  Navarro J.F., Frenk C.S., White S.D.M., 1996, \apj, 462, 563
\reference  Olling R.P., 1995a, AJ, 110, 591
\reference Olling R.P., 1995b, BAAS, 187, 48.05
\reference  Rix H-W., Zaritsky D., 1995, \apj, 447, 82
\reference  Sackett P.D., Rix H-W., Jarvis B.J., Freeman K.C., 1994, \apj, 436, 629
\reference Sackett P.D., Sparke L.S., 1990, \apj, 361, 408
\reference  Sandage A., Tammann G.A., 1981, Revised Shapley-Ames Catalogue of
  Bright Galaxies (Carnegie Inst. of Washington Pub. No. 635)
\reference Schoenmakers R.H.M., Franx M., de Zeeuw P.T., in prep.
\reference  Sicking F.J., PhD-thesis, University of Groningen, 1997
\reference Teuben P., 1991, in Warped Disks and Inclined Rings around Galaxies, (eds. S. Casertano, P.Sackett, F. Briggs), Cambridge Univ. Press.  
\reference Tohline J.E., Simonson G.F., Caldwell N., 1982, \apj, 252, 92
\reference Whitmore B.C., McElroy D., Schweizer F., 1987, \apj, 314, 439
\end{references}
\end{document}